\documentclass[12pt,preprintnumbers]{revtex4-2}
\usepackage{amsfonts}
\usepackage{amsmath}
\usepackage{amssymb}
\usepackage{graphicx}
\usepackage{multirow}
\usepackage[colorlinks=true]{hyperref}
\usepackage{caption}
\usepackage{subcaption}
\newcommand{\al}{\alpha}
\newcommand{\non}{\nonumber}

\newcommand{\re}[1]{(\ref{#1})}
\begin{document}

\title{ClF diatomic molecule: rovibrational spectra}

\date{\today}

\author{Horacio~Olivares-Pil\'on}
\email{horop@xanum.uam.mx}
\affiliation{Departamento de F\'isica, Universidad Aut\'onoma Metropolitana-Iztapalapa,
Apartado Postal 55-534, 09340 M\'exico, D.F., Mexico}

\author{Alexander~V.~Turbiner}
\email{turbiner@nucleares.unam.mx}
\affiliation{Instituto de Ciencias Nucleares, Universidad Nacional
Aut\'onoma de M\'exico, Apartado Postal 70-543, 04510 M\'exico,
D.F., Mexico}

\begin{abstract}
Following the first principles the analytic Born-Oppenheimer (BO) potential
curve for the ground state $X^1\Sigma^+$ of the molecule ClF is proposed for whole range
of internuclear distances $R \in [0,\infty)$.
It is based on matching the perturbation theory at small internuclear distances $R$
and multipole expansion at large distances $R$, it has the form of two-point Pade approximant
and provides 3-4 figures in rovibrational energies. It supports 5719 rovibrational states
with maximal vibrational number $\nu_{max} = 47$ and maximal angular momentum $L_{max} = 210$
including 36 weakly-bound states close to threshold (to dissociation limit) with the energies
$\lesssim 10^{-4}$\,hartree. The van der Waals constant $C^{\rm (ClF)}_6\ \sim\ 29.3$\,a.u. is predicted.

\end{abstract}

\maketitle

\maketitle

\section{Introduction}
The chlorine monofluoride ClF, a heteronuclear diatomic molecule composed of two halogens,
Cl and F, was studied experimentally for the first time in 1942~\cite{AW:1942}.  32 years
later in 1974,  the first theoretical work on this molecule was published~\cite{SL:1974}.
Since then, the  molecule ClF has been the subject of numerous experimental and theoretical
investigations (see e.g.~\cite{DBP:1988,MD:1981,ALBK:2000,VKM:2014} and references
therein). This system is among of diatomic interhalogens, which since long ago is of interest
of laser physics~\cite{DKRK:1982,DPZ:1986} with a chance to lead to a new type of gas lasers.
At room temperature, ClF is a colorless gas, which is highly reactive and corrosive~\cite{GE:1998}.
Needless to say that the presence of the 26 electrons and two heavy nuclei makes chlorine monofluoride
molecule an extremely complex object to study theoretically in traditional approaches. Recently,
it was proposed a new approach based on matching perturbation theory at small distances and
multipole expansion at large distances \cite{Turbiner:2019}, which does not rely heavily on massive
{\it ab initio} calculations. In this approach the nucleus play the role of {\it probs} and
BO potential curve has a meaning of screening function of Coulomb interaction in electronic media.
{  In the past it was successfully checked for H$_2^+$, HeH and H$_2$ diatomics
\cite{OT:2016,OT:2018}, while in \cite{Turbiner:2019} it is illustrated by studying
He$_2^+$ and LiH.}

This Note will be dedicated to study the ground state BO potential curve of ClF molecule
for all internuclear distances.
It is assumed that the first four figures in energy are correction-free: they are not
influenced by adiabatic,
also relativistic and QED corrections and any other corrections like it is for
Helium-like and Lithium-like ionic sequences in atomic physics \cite{Turbiner:2018}.

\section{Potential energy curve}

The chlorine monofluoride ClF is composed of a Chlorine atom Cl ($Z_{\rm Cl}=17$) and
a Fluorine atom F ($Z_{\rm F}=9$). The dissociation energy $E_d$ is related with total energy
$E(R)$ as
\begin{equation}
 E_d\ =\ E(R)\ -\ (E_{\rm Cl}+E_{\rm F})\ ,
\end{equation}
where the ground state energies $E_{\rm Cl}=-460.148$~hartree and $E_{\rm F}=-99.7339$~hartree
\cite{CGD:1993} are taken. The united atom limit of the ClF molecule corresponds to the Iron atom Fe,
where the ground state energy is equal to $E_{\rm Fe}=-1242.985$~hartree \cite{CGD:1993},
it can be considered as a ``splitting" of $(fe)$ nucleus to $(cl)$ plus $(f)$ nuclei.

For small internuclear distances $R\rightarrow 0$  the dissociation energy behaves as
\begin{equation}
\label{EsmallR}
  E_d\ =\ \frac{153}{R}\ +\ \varepsilon_0\ +\ 0 \cdot R\ +\ O(R^2)\ ,
\end{equation}
where the first term is the Coulomb repulsion $Z_{\rm Cl}Z_{\rm F}/R$ and
$\varepsilon_0=E_{\rm Fe} +|E_{\rm Cl}+E_{\rm F} |=-683.1031$~hartree.
Following Bingel~\cite{B:1958} the linear term in $R$ is absent. On the other hand,
the behavior of the dissociation energy at large internuclear distances
$R\rightarrow \infty$  is
\begin{equation}
\label{ElargeR}
  E_d\ =\ -\frac{C_6}{R^6}\ +\ {  \frac{0}{R^7}}\ +\ \frac{C_8}{R^8}\ +\ \cdots\ ,
\end{equation}
cf. \cite{Pauling:1935}, see Eq.(9) therein, for H$_2$ molecule.
{  Let us emphasize that the term $1/R^7$ is absent.} This corresponds to multipole expansion.
The van der Waals constant $C_6$ for ClF molecule, to the best of our knowledge, is not known,
its value can be estimated as
\begin{equation}
\label{c6approx}
    C^{\rm (ClF)}_6\ =\ \frac{2\,C_6^{\rm F}\,C_6^{\rm Cl}}{\frac{\al_{\rm Cl}}{\al_{\rm F}}\, C_6^{\rm F}+\frac{\al_{\rm F}}{\al_{\rm Cl}} \, C_6^{\rm Cl}}\ ,
\end{equation}
see \cite{KTT:1969}, where $\al_{\rm F}=3.74$\,a.u. and $\al_{\rm Cl}=14.6$\,a.u. are the dipole
polarizabilities~\cite{SN:2019} and $C_6^{\rm F}=9.52$\,a.u. and
$C_6^{\rm Cl}=94.6$\,a.u. are the van der Waals constants~\cite{CD:2004} for
the interaction between neutral Fluorine and Chlorine atoms, F$_2$ and Cl$_2$, respectively.
After substituting the respective values into~\re{c6approx}, the final result for the van der Waals constant is
\[
        C^{\rm (ClF)}_6\ \sim\ 29.3\,{\rm a.u.}
\]
In order to construct an analytic expression which interpolates the small $R$~\re{EsmallR} and
large $R$~\re{ElargeR} asymptotic expansions, a rational function in the form of two-point Pad\'e approximant
${\rm Pade}[N/N+5](R)$ is proposed \cite{Turbiner:2019}, so that the potential energy curve is

\[
     E_d(R)\ =\ \frac{1}{R}\,{\rm Pade}[N /N + 5](R)_{n_0,n_{\infty}}\ ,
\]
where $n_0$ is the number of coefficients in the expansion (\ref{EsmallR}) which are reproduced exactly,
while $n_{\infty}$ is the number of coefficients which are reproduced exactly in the expansion at $R=\infty$
at (\ref{ElargeR}). In concrete calculations we choose $N=4$ and $n_0=3, n_{\infty}=2$.
Eventually, we arrive at
\begin{equation}
\label{fitP49}
    E_d(R)\ =\
    \frac{1}{R}\,\frac{153 +\, \sum_{i=1}^{3}a_i R^i-a_4 R^4}{1+\al_1 R+ \al_2 R^2 +\,
    \sum_{i=3}^{7}b_i R^i - \al_3 R^8+b_{9} R^9} \ ,
\end{equation}
with four constrains imposed
\begin{eqnarray}
\label{par-1}
     \al_1 & =  & (a_1-\varepsilon_0)/153\ ,\non \\
     \al_2 & =  & (\varepsilon_0^2 + 153\, a_2  - a_1\varepsilon_0)/153^2\ ,\non \\
     \al_3 & =  & a_3/C_6\ ,\non \\
       b_9 & =  & a_4/C_6\ .
\end{eqnarray}
{  These constraints occur in order to guarantee that in the expansions
of the two-point Pad\'e approximant (\ref{fitP49}) at small and large $R$ the exact
coefficients in front of the first three terms $R^{-1}$, $R^0$ and $R$
at small internuclear distances~\re{EsmallR} and the coefficients in front of
the first two terms $R^{-6}$ and $R^{-7}$ for large internuclear distances~\re{ElargeR}
are reproduced.} Remaining nine parameters in (\ref{fitP49})
are free. These parameters are fixed by choosing nine points among RKR experimental data
~\cite{MD:1981} in domain $R \in [2.6, 3.8]$\,bohr to describe them exactly
\begin{equation}
\begin{array}{lrrr}
 a_1 = & -24.4199 \ ,     & \hspace{0.5cm}b_3 =& -5.78426 \ ,\\
 a_2 = & -30.0557 \ ,     & b_4 = & -2.93344\ ,\\
 a_3 = &  7.75224 \ ,     & b_5 = &  4.05190\ ,\\
 a_4 = &  0.70789 \ ,     & b_6 = & -2.64498\ ,\\
           &   	          & b_7 = &  1.16409\ .
\end{array}
\label{par-2}
\end{equation}

\begin{figure}[h!]
\includegraphics[scale=2.0]{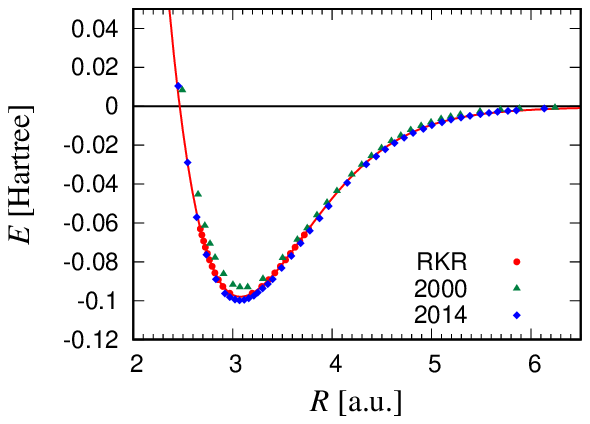}
\caption{BO potential energy (in hartree) for the ground state $X^1\Sigma^+$ of the
   ClF molecule {\it vs} internuclear distance $R$: $(i)$
   the Pad\'e approximant~\re{fitP49} (red line), $(ii)$ RKR experimental
   data~\cite{MD:1981} (red dots) and $(iii)$ points extracted from
   plots of numerical calculations: 2000~\cite{ALBK:2000} (green triangles) and
   2014~\cite{VKM:2014} (blue diamonds). }
\label{FVpotClF}
\end{figure}

Table~\ref{TcompV} presents the dissociation energy $E_d$ obtained from the analytic
expression~\re{fitP49} with parameters (\ref{par-1})-(\ref{par-2}) for different
internuclear distances $R$, which are compared with the all RKR results from
\cite{MD:1981}.
As can be seen that, except for a few points, not less than 4 decimal digits (d.d.)
are in agreement. The function $E_d(R)$~\re{fitP49} is plotted in Fig.~\ref{FVpotClF}
together with the RKR experimental points~\cite{MD:1981} as well as the
numerical theoretical results, extracted from figures in \cite{ALBK:2000} and \cite{VKM:2014}.
It must be emphasized that the results \cite{ALBK:2000} and \cite{VKM:2014},
obtained in different {\it ab initio} calculations, disagree essentially with each
other as well as with RKR experimental data from \cite{MD:1981}
even though all of them agree in predicting the equilibrium distance $R_{eq}$.

The minimum of the BO potential energy curve $E_d(R)$ is calculated by taking the derivative
of~\re{fitP49} $dE_d(R)/R=0$, obtaining  $R_{min}=3.0781$\,bohr and
$E_{min}=-0.09797$\,hartree in a reasonable agreement with the RKR experimental values
$R=3.0771$\,bohr~\cite{WK:1980} and $E_{min}=-0.09796$\,hartree~\cite{MD:1981}. Note that the dissociation energy vanishes, $E_d(R_0)=0$, at $R_0=2.4671$\,bohr.

\begin{table}[h!]
\caption{Dissociation energy (in hartree) for the ground state $X^1\Sigma^+$ for the chlorine
monofluoride molecule ClF as a function of the internuclear distance $R$. First and
second columns are from~\cite{MD:1981}. The last column is obtained from the
approximation~\re{fitP49}.}
\begin{center}
\begin{tabular}{c | cc}
\hline\hline
$R$      & $E_d$~\cite{MD:1981} & $E_d $ fit~\re{fitP49} \\
\hline
2.6743404& -0.06306& -0.06299\,\\
2.6900251& -0.06615& -0.06615\,\\
2.7068437& -0.06929& -0.06933\,\\
2.7249851& -0.07248& -0.07252\,\\
2.7450162& -0.07573& -0.07577\,\\
2.7671260& -0.07902& -0.07904\,\\
2.7922593& -0.08236& -0.08237\,\\
2.8211721& -0.08574& -0.08572\,\\
2.8561321& -0.08918& -0.08914\,\\
2.9012965& -0.09266& -0.09262\,\\
2.9714054& -0.09618& -0.09615\,\\
3.0771185& -0.09796& -0.09797\,\\
3.1955269& -0.09618& -0.09621\,\\
3.2915250& -0.09266& -0.09267\,\\
3.3625787& -0.08918& -0.08918\,\\
3.4238058& -0.08574& -0.08574\,\\
3.4793637& -0.08236& -0.08234\,\\
3.5313312& -0.07902& -0.07900\,\\
3.5806531& -0.07573& -0.07571\,\\
3.6282742& -0.07248& -0.07247\,\\
3.6745724& -0.06929& -0.06928\,\\
3.7197369& -0.06615& -0.06615\,\\
3.7641455& -0.06306& -0.06308\,\\
\hline\hline
\end{tabular}
\end{center}
\label{TcompV}
\end{table}

%--------------------------------------------------------------------------------------------------------------
\section{Rovibrational spectra}

In the Born-Oppenheimer approximation of zero order (when adiabatic correction is neglected),
the rovibrational spectra are calculated by solving the nuclear Schr\"odinger equation
\begin{equation}
\label{NucSc}
\left[-\frac{1}{\mu}\frac{d^2}{dR^2}\ +\ \frac{L(L+1)}{\mu R^2}\ +\ V(R)\right]\,\phi(R)\
=\  E_{\nu,L}\, \phi(R)\ ,
\end{equation}
where $\mu=m_{\rm Cl} m_{\rm F}/(m_{\rm Cl}+ m_{\rm F})$ is the reduced mass
for chlorine and fluorine nuclei, $L$ is the angular momentum
and $E_{\nu,L}$ is the rovibrational energy of the state with vibrational and rotational quantum
numbers $\nu$ and $L$, respectively. The nuclear potential $V(R)$ is the electronic
energy given by {  the two-point Pad\'e approximant} \re{fitP49}.
Equation~\re{NucSc} is solved by using the Lagrange Mesh Method~\cite{DB:2015}
(see for code the Appendix).
Nuclear masses are chosen as $m_{\rm F} =34622.9705$ for $^{19}$F
and $m_{\rm Cl}=63727.3189$ for $^{35}$Cl~\cite{AWT:2003}.

The obtained vibrational spectra $E_{\nu,0}$ is presented in Table~\ref{TvSClF}. Comparison
with the 10 vibrational states reported in~\cite{MD:1981} shows a correspondence in
4 d.d. In total there are 48 vibrational states ($\nu=0,\cdots, 47$).
However, energies of the states with $\nu=45,46,47$ are near the dissociation threshold,
they are smaller than $10^{-4}$\,hartree, thus, beyond our accuracy and have to be excluded.

\begin{table}[h!]
\caption{Vibrational energies $E_{\nu,0}$ in hartree for the ground state $X^1\Sigma^+$
  of the ClF molecule. RKR vibrational states~\cite{MD:1981} for
  $\nu=0,\cdots,10$ are presented in the third column.}
\begin{center}
\scalebox{1}{
\begin{tabular}{r  cc |cc|cc|cc|cc}
\hline\hline
$\nu$&$E_{\nu,0}$ &\cite{MD:1981} &$\nu$&$E_{\nu,0}$&$\nu$&$E_{\nu,0}$&$\nu$&$E_{\nu,0}$&$\nu$&$E_{\nu,0}$\\
\hline
  0& -0.09619& -0.096182&     &         &    &         &    &         &    &         \\
  1& -0.09266& -0.092658&  11 & -0.06002& 21 & -0.03267& 31 & -0.01209& 41 & -0.00117\\
  2& -0.08918& -0.089178&  12 & -0.05703& 22 & -0.03027& 32 & -0.01050& 42 & -0.00074\\
  3& -0.08574& -0.085745&  13 & -0.05409& 23 & -0.02794& 33 & -0.00901& 43 & -0.00042\\
  4& -0.08235& -0.082358&  14 & -0.05121& 24 & -0.02569& 34 & -0.00762& 44 & -0.00021\\
  5& -0.07901& -0.079018&  15 & -0.04839& 25 & -0.02350& 35 & -0.00635& 45 & -0.00009\\
  6& -0.07572& -0.075727&  16 & -0.04562& 26 & -0.02139& 36 & -0.00518& 46 & -0.00003\\
  7& -0.07248& -0.072485&  17 & -0.04291& 27 & -0.01936& 37 & -0.00414& 47 &-0.000003\\
  8& -0.06929& -0.069293&  18 & -0.04026& 28 & -0.01741& 38 & -0.00321&    &         \\
  9& -0.06615& -0.066151&  19 & -0.03766& 29 & -0.01555& 39 & -0.00241&    &         \\
 10& -0.06306& -0.063061&  20 & -0.03514& 30 & -0.01377& 40 & -0.00173&    &         \\
\hline\hline
\end{tabular}}
\end{center}
\label{TvSClF}
\end{table}

\begin{figure}[h!]
\includegraphics[scale=2.0]{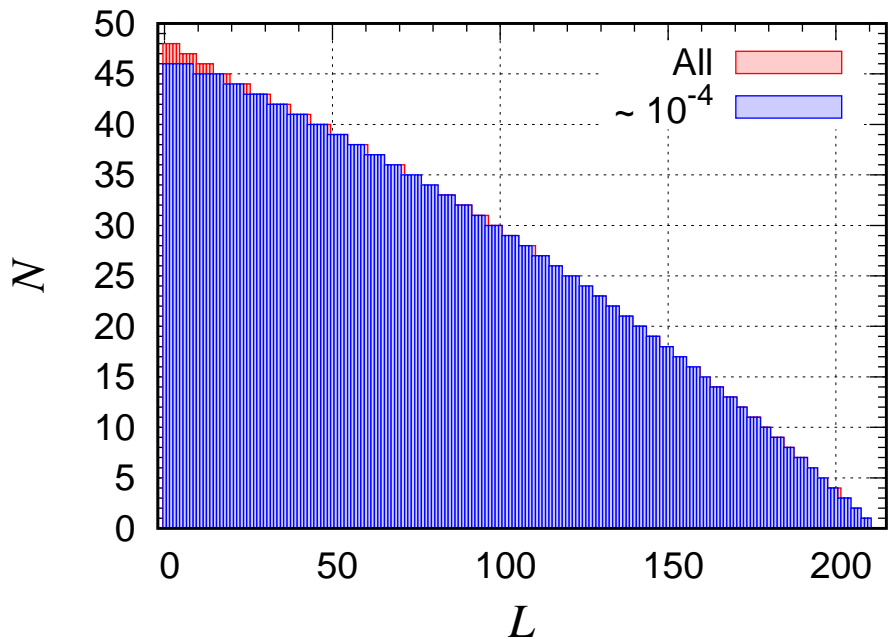}
\caption{Rovibrational spectra for the ground state $X^1\Sigma^+$ of the Chlorine
Monofluoride molecule.  All rovibrational energies are found with absolute accuracy
$\sim 10^{-4}$ hartree. The weakly-bound states of energies $\lesssim 10^{-4}$
hartree are shown in light red.}
\label{rvsClF}
\end{figure}

The complete rovibrational spectra is depicted in the histogram in Figure~\ref{rvsClF}.
In total, there are 5719 rovibrational states with $\nu_{max} = 47$ and  $L_{max} = 210$,
however, as for pure vibrational states $L=0$ the energies of some of them are beyond 
of our accuracy.
Those 5685 rovibrational states obtained with absolute accuracy $\lesssim 10^{-4}$\,hartree
with energies $\lesssim 10^{-4}$\,hartree are indicated in blue in the histogram.
All these 5719 rovibrational energies are obtained by solving the radial
Schr\"odinger equation  (\ref{NucSc}) with the potential (\ref{fitP49})
using the Lagrange-Mesh Method \cite{DB:2015} in a single run for given $L$.
The code and computational details are presented in Appendix.

As a conclusion we have to state that for the ground state of the diatomic molecule ClF
the simple analytic function ~\re{fitP49} with parameters (\ref{par-1})-(\ref{par-2})
allows to describe the all available RKR experimental data in domain $3.8 > R > 2.7$\,bohr
with accuracy of four figures and restore the BO potential curve for all internuclear distances.
All 5719 rovibrational energies with maximal vibrational number $\nu_{max} = 47$ and maximal
angular momentum $L_{max} = 210$ are predicted for the first time.
Making an evident modification of the expansions~\re{EsmallR} and ~\re{ElargeR} accordingly
an analogue of the function ~\re{fitP49} can be written for BO potential curve for
the excited states of ClF.

A.V.T. thanks PASPA-UNAM for a support during his sabbatical stay at University of Miami.

\newpage
\appendix*
\section{}
\label{LMMprogram}

The code based on the Lagrange Mesh Method written in Mathematica-12 is shown by taking the zero angular momentum 
(vibrational states) as the example. 
Mesh points correspond to the zeroes of the Laguerre polynomials. Number of used mesh points varies from 400 
for small $L$ to 800 for largest angular momentum $L$ denoted as $ll$. Scaling parameter $h$, see \cite{DB:2015}, 
used to accelerate convergence, varies for 0.04 to 0.5 depending on the angular momentum $L$. 
For given $L$ the rovibrational spectra is found in a single run. It takes $\sim 10^2$ seconds of CPU time 
on a standard laptop. For confidence the obtained results were recalculated using the code written in FORTRAN-90.

\begin{figure}[h!]
\includegraphics[scale=0.78]{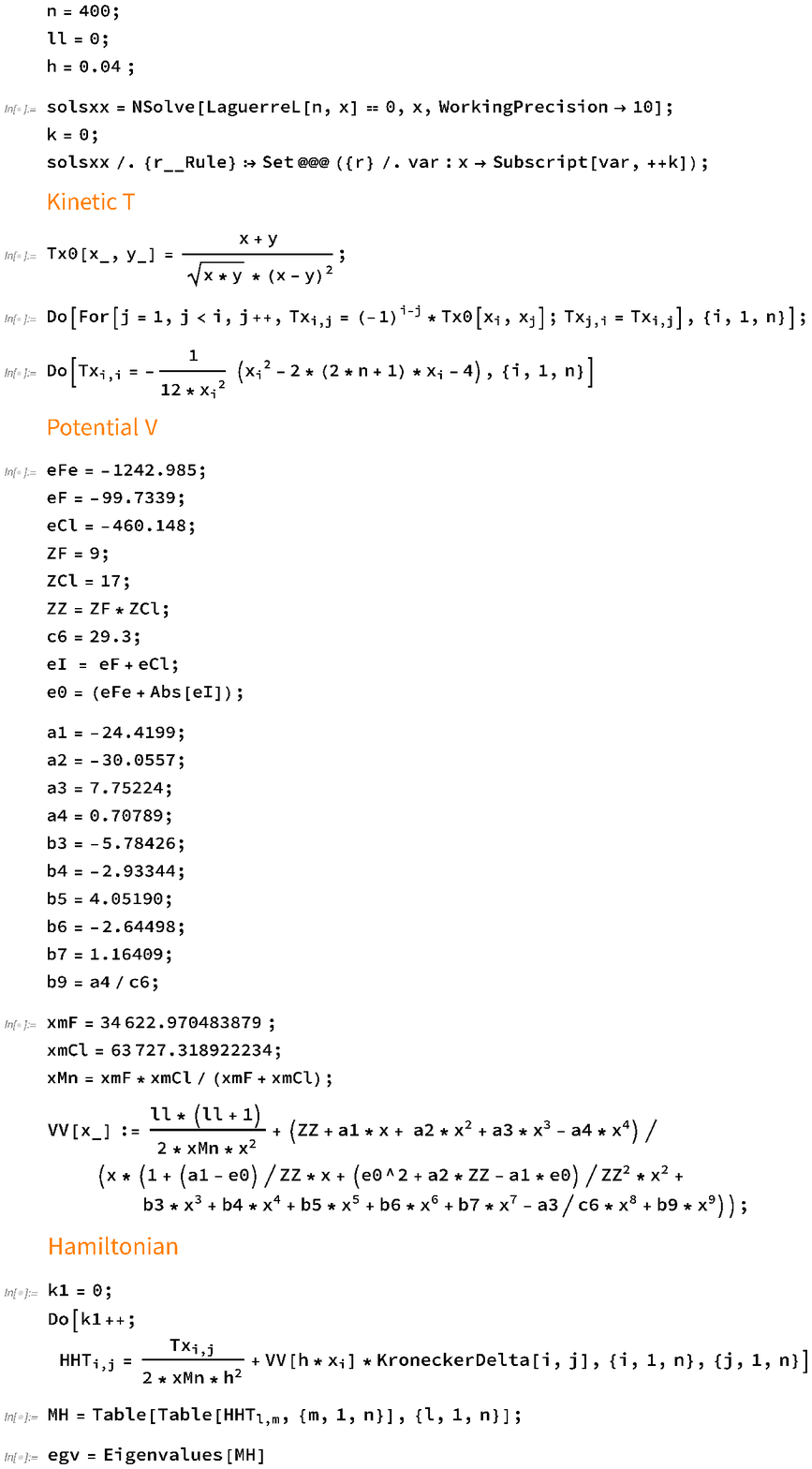}
\end{figure}

\end{document}